\pdfoutput=1
\documentclass[11pt,a4paper]{article}
\usepackage{fullpage}
\usepackage{authblk}
\usepackage{amsmath,amsthm,amsfonts}
\usepackage{thm-restate}
\usepackage{bbm}
\usepackage{xspace}
\usepackage{tikz}
\usepackage{fullpage}
\usepackage{newpxtext,newpxmath}
\usepackage{color}
\usepackage{hyperref}

\definecolor{blue}{RGB}{0,50,200}
\definecolor{magenta}{RGB}{255,0,255}
\hypersetup{colorlinks,linkcolor=blue,urlcolor=blue,citecolor=magenta}

\usetikzlibrary{positioning}

\newtheorem{theorem}{Theorem}

\newtheorem{lemma}[theorem]{Lemma}
\newtheorem{observation}[theorem]{Observation}

\newcommand{\leftof}{\textrm{---} \xspace}
\newcommand{\among}[2]{\left( \substack{#1\\#2} \right)}

\newcommand{\rand}{\textsc{Rand}\xspace}
\newcommand{\OPT}{\textsc{Opt}\xspace}
\newcommand{\DET}{\textsc{Det}\xspace}
\newcommand{\ONL}{\textsc{Onl}\xspace}
\newcommand{\OFF}{\textsc{Off}\xspace}
\newcommand{\ALG}{\textsc{Alg}\xspace}

\title{Learning Minimum Linear Arrangement\\of Cliques and Lines\thanks{%
This paper has been supported by Polish National Science Centre grant
2022/45/B/ST6/00559 and the Austrian Science Fund (FWF) project I 5025-N
(DELTA). For the purpose of Open Access, the authors have applied CC-BY public
copyright license to any Author Accepted Manuscript (AAM) version arising from
this submission.
}}

\begin{document}

\author[1]{Julien Dallot}
\author[1]{Maciej Pacut}
\author[2]{Marcin Bienkowski}
\author[1]{Darya Melnyk}
\author[1]{Stefan Schmid}
\affil[1]{Technische Universität Berlin, Germany}
\affil[2]{University of Wroclaw, Poland}
\date{}

\maketitle

\begin{abstract}
In the well-known Minimum Linear Arrangement problem (MinLA), the goal is to
arrange the nodes of an undirected graph into a permutation so that the total
stretch of the edges is minimized. This paper studies an online (learning)
variant of MinLA where the graph is not given at the beginning, but rather
revealed piece-by-piece. The algorithm starts in a fixed initial permutation,
and after a piece of the graph is revealed, the algorithm must update its
current permutation to be a MinLA of the subgraph revealed so far. The objective
is to minimize the total number of swaps of adjacent nodes as the algorithm
updates the permutation.

The main result of this paper is an online randomized algorithm that solves this
online variant for the restricted cases where the revealed graph is either 
a~collection of cliques or a~collection of lines. We show that the algorithm is
$8 \ln n$-competitive, where $n$ is the number of nodes of the graph. We
complement this result by constructing 
an~asymptotically matching lower bound of $\Omega(\ln n)$.
\end{abstract}

\section{Introduction}

Minimum Linear Arrangement
(MinLA)~\cite{RaoR04,treeMLA,hypercubeMLA,hypercubeMLA,Jordi_MinLA}
is a classic combinatorial optimization problem that arises in the field of
graph theory and network design. The objective is to arrange the nodes of an
undirected graph along a~straight line in such a way that the sum of the
distances between adjacent nodes is minimized. The problem finds applications in
various domains, including VLSI circuit design, transportation network planning,
and communication network layout~\cite{Harper, Sartaj, AvinDPS23}. The problem
is formally defined as follows. Let $G = (V, E)$ be a graph with $n$ nodes and
$m$ edges. The goal of the MinLA problem is to find a
permutation $\pi$ of the nodes of $G$ so that the total distance between the
edges is minimized. Formally, it amounts to finding $\pi \in S_n$ such that the
quantity
\begin{align*}
  \sum\limits_{(x,y) \in E} |\pi(x) - \pi(y)|
\end{align*}
is minimized, where $\pi(x)$ denotes the position of node $x$ in the permutation $\pi$.

In this paper, we study the online \emph{learning} variant of the MinLA problem. In
the online learning variant, the graph is not given right away but rather
revealed piece-by-piece in a~potentially adversarial order. We call $G_0, G_1,
\dots G_k$ the sequence of the revealed graphs, where $G_0$ is the empty graph,
each $G_i$ is a~super-graph of its predecessors, and $G_k = G$. The online algorithm
starts with a given initial permutation $\pi_0$ and, upon receiving graph $G_i$,
it \emph{must} update its current permutation $\pi_{i-1}$, so that \emph{the new permutation
$\pi_{i}$ is a MinLA of $G_i$}. The~goal of the online MinLA problem is to find
a~sequence~of permutations $\pi_0, \dots, \pi_k$, such that the total cost of the
permutation updates is minimized. More formally, the algorithm
must minimize the quantity
\begin{align*}
  \sum\limits_{i=0}^{k-1} d(\pi_{i}, \pi_{i+1}),
\end{align*}
where $d$ is the cost to update a permutation. We choose $d$ to be Kendall's
$\tau$ distance, that is, the minimum number of pairwise swaps of adjacent elements 
required to change one permutation into the other.

We use the notion of competitive ratio to assess the performance of our online
algorithms~\cite{borodin-book}. For a given request sequence $\sigma$, let
$\ONL(\sigma)$ and $\OPT(\sigma)$ be the cost incurred by an online algorithm
$\ONL$ and an optimal offline algorithm $\OPT$, respectively. In contrast to
$\ONL$, $\OPT$ knows the entire request sequence $\sigma$ ahead of time. The goal
is to design online algorithms for online learning MinLA that provide worst-case
guarantees. In particular, $\ONL$ is said to be $c$-competitive if for any input
sequence $\sigma$ it holds that $\ONL(\sigma) \le c\cdot \OPT(\sigma)$.

This work aligns with a broader line of work that searches for efficient online
algorithms for classic offline optimization problems. As dealing with an unknown
future is a ubiquitous scenario, there exists an extensive literature on online
variants of well-known problems, including set
cover~\cite{AlonAABN09,BuchbinderN09}, graph matching \cite{Mehta_Matching,
Gamlath_Matching}, graph coloring \cite{VISHWANATHAN_ColoringRandomized,
Epstein_Coloring}, load balancing \cite{PhillipsW93_LoadBalancing}, maximum
independent set \cite{HALLDORSSON_MIS}, and bin packing \cite{Seiden_BinPacking,
Balogh_BinPacking}. So far, the MinLA problem has received considerable
attention in the offline setting~\cite{treeMLA,hypercubeMLA,Eikel_MinLA,Jordi_MinLA} 
either due to its many application
cases~\cite{Harper, Sartaj, AvinDPS23} or its theoretical
importance~\cite{Ambuhl_MinLA, Even_MinLA}, and developing a competitive online
algorithm is a natural step in this line of research.

Our work is related the closest to work of Olver et al.~\cite{ILU}, who
introduced the \emph{dynamic} MinLA problem and its
variants (such as the itinerant list update problem). Their setting is less
restricted than ours and it leaves more freedom on how to serve a request: when
an edge is revealed in their variant, an algorithm simply pays the distance
between its endpoints in its current permutation, and then may optionally rearrange
the nodes to save cost on future requests, the goal being to minimize the total
incurred cost. In particular, in their dynamic setting an~algorithm does not have to
maintain a MinLA of the graph induced by edges seen so far while our setting
is learning the graph and remembers previous edges.

Their problem received the most attention in its offline setting: Olver et
al.~\cite{ILU} presented a randomized $(\log^{2}n)$-approximation algorithm,
later improved by Bienkowski and Even to a~deterministic algorithm with a $(\log
n \log \log n)$-approximation~\cite{BienkowskiE24}. While the online variant of
the dynamic MinLA problem is regularly mentioned~\cite{ILU,BienkowskiE24}, 
no advances were made on this front; the only results so far were a lower bound of $\Omega(\log
n)$ for the competitive ratio of any randomized online algorithm~\cite{ILU}, and
a trivial upper bound of $O(n)$ held by the algorithm that never changes its
permutation. 

To the best of our knowledge, this paper hence is the first to tackle a
variant of the problem in the online setting hopefully paving the way
towards more general results for the online dynamic MinLA problem.

\subsection{Contributions}

We consider online learning MinLA problem in a restricted case where each
graph $G_i$ in the input is either a collection of disjoint cliques or a collection of
disjoint lines (paths). This restriction captures fundamental questions that arise in
a general setting: (1) where to move the two disjoint components that have just been
connected by revealing of a new piece of the graph, and (2) which orientation
should the connected component have in the permutation.

We develop two online algorithms for this setting: a deterministic one and a
randomized one. We begin with the deterministic online algorithm, prove it is
$2n$-competitive and show that the analysis is tight. The main contribution of
this paper, however, is the randomized online algorithm that is $8 \ln n$-competitive
against the oblivious adversary~\cite{borodin-book}. Finally, we show a randomized lower
bound of $\Omega(\log n)$, proving that our algorithm is asymptotically optimal.

\subsection{Motivation: Dynamic Virtual Network Embedding}
\label{ssec:motiv}

With the increasing popularity of data-centric workloads, such as distributed
machine learning and scale-out databases, datacenters are witnessing a
significant surge in network traffic~\cite{roy2015inside}. The performance of
these distributed applications heavily relies on the efficiency of the
underlying communication networks~\cite{mogul2012we}.

An intriguing approach to improve the efficiency of communication networks is to
adjust the network in a demand-aware and online manner. This can be done by
leveraging resource allocation flexibilities provided by virtualization. When
communication requests arrive over time, algorithms can decide to move
frequently communicating nodes topologically closer. Since relocation comes at a
cost, the task is to strike a~balance between the benefits and the costs of the
adjustments. The underlying optimization problem involves dynamically embedding
a~virtual network on a physical network topology.

The basic virtual network embedding problem of finding a mapping from a request
graph to the virtual network is
well-studied~\cite{FischerBBMH13,Rost018,ChowdhuryRB12,Diaz02, Rost019}, and
many problem variants have been shown to be
$\mathcal{NP}$-complete~\cite{FischerBBMH13,Rost018,ChowdhuryRB12}. One of the
most studied problems in graph embedding is the minimum linear arrangement
problem. This problem is $\mathcal{NP}$-hard in general, therefore, significant efforts have
been made to solve it for restricted guest graphs such as trees \cite{treeMLA}
or incomplete hypercubes \cite{hypercubeMLA}.

The problem variant of \emph{dynamically} embedding a virtual network on a
physical network topology is, however, more challenging. So far, the dynamic
problem has been considered for two fundamental physical network topologies: (1)
a set of capacitated clusters connected in a~clique and (2) the line topology.
In the first setting, the problem is known as the \emph{online graph
partitioning} problem~\cite{Avin2020,Henzinger0019,HenzingerNRS21,PacutP020},
and in the latter setting, it is known as the \emph{dynamic minimum linear
arrangement} problem (a.k.a. itinerant list update problem~\cite{ILU}).

The online learning MinLA model is inspired by the learning variant of online graph
partitioning, introduced at SIGMETRICS 2019~\cite{Henzinger0019}, with followup
work at INFOCOM 2020~\cite{PacutP020} and SODA 2021~\cite{HenzingerNRS21}. The
goal is to learn and embed the communication pattern (the guest graph) over time
in an online manner. 

We develop an \emph{optimally competitive} randomized online algorithms for
online learning MinLA for two fundamental traffic patterns: collection of lines and
collection of cliques. These topologies represent the most sparse and most dense
traffic patterns and are an important stepping stone toward more complex network
topologies.

\subsection{Related Work}

The model introduced in this paper is related to two online problems:
\emph{online graph partitioning}, introduced at DISC 2016~\cite{Avin2020}, and
\emph{dynamic minimum linear arrangement}, introduced at WAOA 2018~\cite{ILU}.

In the online graph partitioning problem~\cite{Avin2020}, a network topology
connecting $\ell$ clusters each containing $k$ virtual nodes is considered. The
requests concern pairs of nodes, and if the requested nodes are collocated
within a single cluster, the request costs $0$, and otherwise, it costs $1$. For
this problem, the best known competitive ratio is $(k\cdot \ell \cdot
2^{O(k)})$-competitive~\cite{Marcin2021}, and no deterministic algorithm can be
better than $\Omega(k\cdot \ell)$-competitive~\cite{PacutP020}. Significant
research efforts on the graph partitioning problem have focused on the learning
variant~\cite{Henzinger0019,PacutP020,HenzingerNRS21}.

In the dynamic minimum linear arrangement problem~\cite{ILU}, the underlying
topology is a~line and the requests are pairs of virtual nodes. The cost for
serving a request is proportional to the distance between the corresponding
virtual nodes, but collocation is not strictly enforced. A $(\log n \cdot \log
\log n)$-approximation algorithm for this problem has been derived in the
offline setting, where all the requests are known in
advance~\cite{ILU,BienkowskiE24}. The paper of~\cite{ILU} shows 
a~randomized lower-bound of
$\Omega(\log n)$ for the dynamic MinLA problem in the
online setting. So far, the only general upper bound is a trivial $O(n)$ bound, 
which can be achieved
even by an algorithm that never migrates (however we point out that in the model
studied in this paper, applying this algorithm is not viable). A simple online
algorithm that moves the smaller component towards the larger has an upper bound
of $O(n^2\cdot \log n)$ on the total incurred cost~\cite{AvinDPS23}, but under
strict competitive analysis this algorithm is $\Omega(n)$-competitive, and
neither the algorithm nor the analysis translates to our setting.

\subsection{Organization}

This manuscript is organized as follows. In Section~\ref{sec:det}, we design and
analyze a simple deterministic algorithm that attains a linear competitive ratio
for the case of embedding a~collection of lines or a collection of cliques. In
Section~\ref{sec:rand_for_cliques}, we design a randomized algorithm attaining 
a~logarithmic competitive ratio for the case of embedding a collection of cliques.
In Section~\ref{sec:rand_for_lines}, we adapt the previous algorithm to attain a
logarithmic competitive ratio for the case of embedding a collection of lines.
In Section~\ref{sec:lbs}, we prove the tightness of these algorithms. We
conclude in Section~\ref{sec:conclusions}.

\section{A Deterministic Algorithm}
\label{sec:det}

In this section, we present a simple $O(n)$-competitive deterministic algorithm
\DET for solving online MinLA for both cases where the subgraphs $G_i$ either
are collections of cliques or collection of lines. In Section~\ref{sec:lbs} we
show that the analysis of \DET algorithm is tight for both graph classes. This
algorithm is an adaptation of an algorithm Perfect Partition Learner for the
\emph{online graph partitioning problem}~\cite{PacutP020}.

The algorithm \DET is defined as follows. Upon each request $G_i$, \DET updates
the permutation to an arbitrary MinLA of $G_i$ that minimizes the distance to
$\pi_{0}$.

In the rest of this section, we argue that the competitive ratio of this
algorithm is linear in terms of the number of nodes.
\begin{theorem}	\label{thm:det-linear}
  \DET is $(2n-2)$-competitive for the case where the graph is either 
  a collection of lines or a~collection of cliques.
\end{theorem}

\begin{proof}
Fix a request sequence $G_0, G_1, \dots G_k$, an initial permutation $\pi_{0}$
and an optimal offline algorithm \OPT with final permutation $\pi_{k}^{\OPT}$.
We denote $d(\pi_{0}, \pi_{k}^{\OPT})$ by $\Delta^*$.

First, we claim that for any permutation $\pi_{i}$ reached by $\DET$ throughout
its execution, we have $d(\pi_{0},\pi_{i}) \leq \Delta^*$. Notice that, whether
the subgraphs are collections of cliques or collections of lines, a permutation
that is a MinLA of $G_k$ is always also a MinLA of any $G_i$ for $i \le k$. As a
result, \DET considers $\pi_k^{\OPT}$ as a potential next permutation each time
a new subgraph is revealed. Since \DET picks the permutation that minimizes its
distance from $\pi_0$, we directly derive that $d(\pi_{0},\pi_{i}) \leq
\Delta^*$.

Second, we claim that the cost of serving each request by \DET is at most
$2\cdot\Delta^*$. We argue as follows. Consider an update from a permutation
$\pi_{i-1}$ to $\pi_{i}$. In the previous paragraph, we argued that both these
permutations are no further than $\Delta^*$ away from $\pi_{0}$. Using the
triangle inequality we have that $d(\pi_{i-1},\pi_{i}) \le
d(\pi_{i-1},\pi_{0}) + d(\pi_{0}, \pi_{i}) \le 2\cdot \Delta^*$.

Finally, we bound the competitive ratio. We have that $\OPT \geq d(\pi_{0},
\pi_{k}^{\OPT}) = \Delta^*$, as both \OPT and \DET start in the same
permutation. The sequence of subgraphs $G_1, \dots G_k$ has length at most
$n-1$. In the previous paragraph, we argued that each request costs at most
$2\cdot \Delta^*$, hence the total cost of \DET is at most $2(n-1) \cdot
\Delta^*$. Hence, $\DET / \OPT \le 2(n-1)\cdot\Delta^* / \Delta^* = 2(n-1)$.
\end{proof}

In Section~\ref{sec:lbs}, we show that our analysis of \DET algorithm is tight.

\section{An Optimal Randomized Algorithm for Cliques}
\label{sec:rand_for_cliques}

\subsection{Description of the Algorithm}

In this section, we present a randomized algorithm \rand for the MinLA problem
when the subgraphs $G_0, \dots G_k$ all are collections of cliques. We will
prove that this algorithm holds an~expected competitive ratio of $4 \ln n$
against the oblivious adversary~\cite{borodin-book}.

Let $X_{i}$ and $Z_{i}$ be the nodes of the two components (cliques) that merge
between $G_{i}$ and $G_{i+1}$.
In response to the reveal of $G_{i+1}$, \rand handles the necessary update by
placing $X_i$ and $Z_i$ next to each other in $\pi_{i+1}$. \rand only considers
two of the potentially many possible permutations where $X_i$ and $Z_i$ end up
next to each other: starting from $\pi_i$, either the nodes of $X_i$ move in
direction of $Z_i$ or conversely, the nodes of $Z_i$ move in direction of $X_i$.

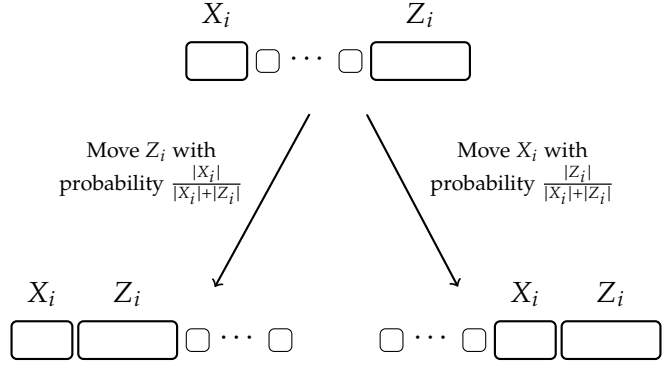
\begin{figure}[t]
  \centering
  \tikzstyle{comp}=[draw, thick, rounded corners=2pt, minimum height = 0.5cm]
  \tikzstyle{minicomp}=[draw, rounded corners=2pt, minimum height = 0.3cm]
  \begin{tikzpicture}[transform shape, scale=1]

  \def\distanceVertical{3.7cm}
  \def\distanceHorizontal{2.3cm}

  \def\widthX{0.8cm}
  \def\widthZ{1.3cm}
  \def\widthY{0.3cm}

  \def\distanceXandY{0.1cm}
  \def\distanceXandZ{0.05cm}
  \def\distanceY{0cm}

  \node[comp, minimum width=\widthX] at (0, 0 - 0*\distanceVertical) (compX1) {};
  \node (compY1) [minicomp, right=0cm and \distanceXandY of compX1, minimum width=\widthY] {};
  \node (compBetween1) [right=0cm and \distanceY of compY1]  {$\cdots$};
  \node (compYprime1) [minicomp, right=0cm and \distanceY of compBetween1, minimum width=\widthY] {};
  \node (compZ1) [comp, minimum width=\widthZ, right=0cm and \distanceXandY of compYprime1] {};

  \node (X1) [above=0.05cm and 0cm of compX1] {$X_{i}$};
  \node (Z1) [above=0.05cm and 0cm of compZ1] {$Z_{i}$};
  
  \node[comp, minimum width=\widthX] at (-\distanceHorizontal, 0 - 1*\distanceVertical) (compX2) {};
  \node (compZ2) [comp, minimum width=\widthZ, right=0cm and \distanceXandZ of compX2] {};
  \node (compY2) [minicomp, right=0cm and \distanceXandY of compZ2, minimum width=\widthY] {};
  \node (compBetween2) [right=0cm and \distanceY of compY2]  {$\cdots$};
  \node (compYprime2) [minicomp, right=0cm and \distanceY of compBetween2, minimum width=\widthY] {};

  \node (X2) [above=0.05cm and 0cm of compX2] {$X_{i}$};
  \node (Z2) [above=0.05cm and 0cm of compZ2] {$Z_{i}$};
  
  \node [minicomp, minimum width=\widthY] at (\distanceHorizontal, 0 - 1*\distanceVertical) (compY3) {};
  \node (compBetween3) [right=0cm and \distanceY of compY3]  {$\cdots$};
  \node (compYprime3) [minicomp, right=0cm and \distanceY of compBetween3, minimum width=\widthY] {};
  \node (compX3) [comp, minimum width=\widthX, right=0cm and \distanceXandY of compYprime3] {};
  \node (compZ3) [comp, minimum width=\widthZ, right=0cm and \distanceXandZ of compX3] {};

  \node (X3) [above=0.05cm and 0cm of compX3] {$X_{i}$};
  \node (Z3) [above=0.05cm and 0cm of compZ3] {$Z_{i}$};

  \draw [thick, shorten <=0.8cm, shorten >=0.8cm, ->] (compYprime1.west) -- node [pos=0.55, anchor=south east] {\normalsize $\substack{ \text{Move } Z_{i} \text{ with}\\ \text{probability } \frac{|X_{i}|}{|X_{i}| + |Z_{i}|}}$} (compY2.west);
  \draw [thick, shorten <=0.8cm, shorten >=0.8cm, ->] (compYprime1.west) -- node [pos=0.55, anchor=south west] {\normalsize $\substack{\text{Move } X_{i} \text{ with}\\ \text{probability } \frac{|Z_{i}|}{|X_{i}| + |Z_{i}|}}$} (compYprime3.east);
  \end{tikzpicture}
  \caption{The two possible actions of \rand to move the merging components $X_{i}$ and $Z_{i}$ together when $G_{i+1}$ is revealed}
  \label{fig:moving_possibilities}
\end{figure}

Our algorithm chooses between those two choices by flipping a biased coin:
$X_{i}$ moves with probability $|Z_{i}| / (|X_{i}| + |Z_{i}|)$ and $Z_{i}$ moves
with probability $|X_{i}| / (|X_{i}| + |Z_{i}|)$. Figure
\ref{fig:moving_possibilities} summarizes the possible actions of our algorithm.

In the rest of this section, we will prove the following theorem.
\begin{restatable}{theorem}{thFinalCompRatioCliques}
  The algorithm \rand is $4 \ln n$-competitive 
  for online MinLA when all the revealed subgraphs are collections of cliques.
  \label{thm:final_comp_ratio_cliques}
\end{restatable}

To prove the theorem, we first introduce necessary notation.

\subsection{Notation}
\label{ssec:notation-cliques}

Let $\mathcal{C}_{i}$ be the set of components of $G_{i}$, a component is the
set of nodes of a connected component. We call $X_{i}, Z_{i} \in
\mathcal{C}_{i}$ the two components that merge into a bigger clique when
$G_{i+1}$ is revealed.

For $i \in \{0 \dots k\}$, let $T, U \in \mathcal{C}_{i}$ be two components. For
these, we define:
\begin{itemize}
  \item $T \times U = \left\{ (t, u) | t \in T, u \in U \right\}$ is the Cartesian product of $T$ and $U$.
  \item $T \leftof U$ is the event that $T$ is on the left of $U$; $P[T \leftof U]$ is the associated probability that \rand is in a configuration with such order between the components.
  \item $L_{\pi_{i}}$ is the set of all node pairs $(x, y)$ such that $x$ is on the left of $y$ in $\pi_{i}$.
  \item $L_{T,U}$ is the set of all pairs that contain exactly one node from $T$ and one node from $U$ in any order, 
    i.e., $L_{T, U} = T \times U \cup U \times T$.
\end{itemize}
Notice that the event $T \leftof U$ is not indexed by $i$. This is because, by
the properties of our algorithm, the relative position of two components $T$ and
$U$ can change only if a node of $T$ or $U$ is requested; but if this happens
then the requested component stops being a component (it becomes a part of a
bigger one) and the event $T \leftof U$ has no more meaning. Hence any two
components will have only one relative position (if they happen to exist at the
same moment) and the event $T \leftof U$ is independent from $i$.

\subsection{Upper Bound on the Expected Cost}
\label{ssec:moving}

We start the analysis with a lemma that gives the probability ruling the relative position of any two components throughout the execution of \rand.

\begin{lemma}
  \label{lem:proba_left_or_right}
  Let $i \in \{0 \dots k\}$, let $X$ and $Y$ be two components of $\mathcal{C}_{i}$.
  It holds that
  \begin{align*}
    P[ X \leftof Y] = \frac{ |X \times Y \cap L_{\pi_{0}}|}{|X| \cdot |Y|}.
  \end{align*}
\end{lemma}

\begin{proof}
  We prove the claim by induction on $i$.
  If $i = 0$ then nothing was revealed so far, the components are the nodes themselves and the claim holds.

  Let $i \in \{1 \dots k\}$, we assume that the claim holds for all components
  when $G_i$ is revealed. We then prove that the claim also holds after
  $G_{i+1}$ was revealed. Let $Y$ be a component different from the merging
  components $X_{i}$ and $Z_{i}$. We compute the probability that $(X_{i} \cup
  Z_{i}) \leftof Y$, i.e., that the newly formed component $X_{i} \cup Z_{i}$ is
  on the left of $Y$ after $X_{i}$ merged with~$Z_{i}$.
  \begin{align*}
    P[(X_{i} \cup Z_{i}) \leftof Y] 
      &= P[X_{i} \leftof Y] \cdot P[Z_{i} \leftof Y]
       + P[X_{i} \leftof Y] \cdot P[Y \leftof Z_{i}] \cdot \frac{|X_{i}|}{|X_{i}| + |Z_{i}|}\\
      & \quad\quad + P[Z_{i}  \leftof Y] \cdot P[Y \leftof X_{i}] \cdot \frac{|Z_{i}|}{|X_{i}| + |Z_{i}|}\\
      &= P[X_{i} \leftof Y ] \cdot \frac{|X_{i}|}{|X_{i}| + |Z_{i}|}
        + P[Z_{i} \leftof Y ] \cdot \frac{|Z_{i}|}{|X_{i}| + |Z_{i}|} \\
      &= \frac{|(X_{i} \cup Z_{i}) \times Y \cap L_{\pi_{0}}|}{|X_{i} \cup Z_{i}| \cdot |Y|}.
  \end{align*}

  We applied the induction hypothesis to obtain the final result.
  By symmetry, we have the same results for computing $P[Y \leftof (X_{i} \cup Z_{i}) ]$.
  The induction holds and the claim follows.
\end{proof}

Lemma~\ref{lem:proba_left_or_right} implies that, at any moment, the probabilistic relative position of any two components of \rand's permutation only depends on the initial permutation after $G_{i}$ was revealed.
In particular, the probability distribution of the permutation $\pi_i$ is independent on the reveal order.

We now use this lemma to upper-bound the cost of an update. For any set of nodes
$U$ and~$T$, we call $M_{U, T}^{i}$ the total number of swaps that occurs
between a node of $U$ and a node of~$T$ as \rand updates its permutation from
$\pi_{i}$ to $\pi_{i+1}$.
\begin{lemma}
  \label{lem:expected_number_of_swaps}
  Let $0 \le i \le k-1$, let $Y \in \mathcal{C}_i$ different from the merging components $X_{i}$ and $Z_{i}$.
  It holds that
  \begin{align*}
      \frac{1}{2} \mathbb{E}[M_{Y, X_{i} \cup Z_{i}}^{i}] 
      & \leq |(L_{\pi_{0}} \setminus L_{\pi_{k}^{\OPT}}) \cap L_{X_{i}, Y}| \cdot \frac{|Z_{i}|}{|X_{i}| + |Z_{i}|} \\
      & \quad\quad + |(L_{\pi_{0}} \setminus L_{\pi_{k}^{\OPT}}) \cap L_{Z_{i}, Y}| \cdot \frac{|X_{i}|}{|X_{i}| + |Z_{i}|}.
  \end{align*}
\end{lemma}

\begin{proof}
  The nodes of $Y$ will swap with nodes of $X_{i}$ or $Z_{i}$ if and only if $Y$
  is between $X_{i}$ and $Z_{i}$ when $G_{i+1}$ is revealed. Expressed in the
  terms of Lemma~\ref{lem:proba_left_or_right} this condition becomes: the nodes
  of~$Y$ will swap with those of $X_{i}$ or $Z_{i}$ only if $X_{i}$ is on the
  left of $Y$ and $Z_{i}$ on the right, or conversely.

  Multiplying the expected cost in case $Y$ crosses one of the merging
  components by the probability that $Y$ is between $X_{i}$ and $Z_{i}$, we
  obtain that
  \[
    \mathbb{E}[M_{Y, X_{i} \cup Z_{i}}^{i}] 
    = 2 |Y| \frac{|Z_{i}| \cdot |X_{i}|}{|X_{i}| + |Z_{i}|} \cdot \left( 
        P[X_{i} \leftof Y] \cdot P[Y \leftof Z_{i}] + P[Z_{i} \leftof Y] \cdot P[Y \leftof X_{i}]
      \right).
  \]

In $\pi_{k}^{\OPT}$, the nodes from $X_{i} \cup Z_{i}$ occupy contiguous places.
Hence, $Y$ is either on the right or on the left of $X_{i} \cup Z_{i}$ in $\pi_{k}^{\OPT}$; using the problem's symmetries we assume that $Y$ is on the left of $X_{i} \cup Z_{i}$ in $\pi_{k}^{\OPT}$ without loss of generality.
We now upper-bound the probabilities $P[Y \leftof X_{i}]$ and $P[Y \leftof Z_{i}]$ in the last equation by $1$ and apply Lemma \ref{lem:proba_left_or_right}:

  \begin{align*}
    \frac{1}{2} \mathbb{E}[M_{Y, X_{i} \cup Z_{i}}^{i}] 
      &\le |Y| \cdot \frac{|Z_{i}| \cdot |X_{i}|}{|X_{i}| + |Z_{i}|} \cdot \left( P[X_{i}-Y] + P[Z_{i} - Y]\right) \\
      & = |X_{i} \times Y \cap L_{\pi_{0}}| \cdot \frac{|Z_{i}|}{|X_{i}| + |Z_{i}|} 
      + |Z_{i} \times Y \cap L_{\pi_{0}}| \cdot \frac{|X_{i}|}{|X_{i}| + |Z_{i}|}.
  \end{align*}
  All node pairs in $X_{i} \times Y \cap L_{\pi_{0}}$ have a node from $X_{i}$ on the left and a node from $Y$ on the right.
  Likewise, all pairs in $Z_{i} \times Y \cap L_{\pi_{0}}$ have a node from $Z_{i}$ on the left part and a node from $Y$ on the right part.
  Hence, neither of those two sets share a pair with $Y \times X_{i}$, $Y \times Z_{i}$ or $\pi_{k}^{\OPT}$.
  As a~result, it holds that
  \begin{align*}
    |X_{i} \times Y \cap L_{\pi_{0}}| = |(L_{\pi_{0}} \setminus L_{\pi_{k}^{\OPT}}) \cap L_{X_{i}, Y}| 
  \end{align*}
  and
  \begin{align*}
    |Z_{i} \times Y \cap L_{\pi_{0}}| = |(L_{\pi_{0}} \setminus L_{\pi_{k}^{\OPT}}) \cap L_{Z_{i}, Y}|,
  \end{align*}
  which ends the proof.
\end{proof}

We state an auxiliary lemma before stating the final theorem.

\begin{restatable}{lemma}{lemHarmSumForMovingCosts}
  Let $s_{1}, s_{2}, \dots s_{N}$ be a series of strictly positive natural numbers, let $S = \sum_{i=1}^{N} s_{i}$ be their sum.
  It holds that
  \begin{align*}
          \sum_{i=1}^{N} \frac{s_{i}}{\sum_{j=1}^{i} s_{j}} \leq H_S.
  \end{align*}
  where $H_S = 1 + \frac{1}{2} + \dots + \frac{1}{S}$ is the harmonic sum.
  \label{lem:harm_sum_for_moving_costs}
\end{restatable}

\begin{proof}
  \[
    \sum_{i=1}^{N} \frac{s_{i}}{\sum_{j=1}^{i} s_{j}} 
    = \sum_{i=1}^{N} \sum_{k=1}^{s_{i}} \frac{1}{\sum_{j=1}^{i} s_{j}}
    \leq \sum_{i=1}^{N} \sum_{k=1}^{s_{i}} \frac{1}{\sum_{j=1}^{i-1} s_{j} + k}
    = \sum_{i=1}^{S} \frac{1}{i}.
  \qedhere
  \]
\end{proof}

\begin{theorem}
  \label{th:ub_moving_cost}
  Let $M$ be the random variable equal to the cost of \rand. Then,
  $\mathbb{E}[M] \le 4 H_n \cdot |L_{\pi_{0}} \setminus L_{\pi_{k}^{\OPT}}|$,
  where $H_n$ is the harmonic sum.
\end{theorem}

\begin{proof}[Proof of Theorem \ref{th:ub_moving_cost}]
We use Lemma~\ref{lem:expected_number_of_swaps} to upper-bound $M$.
It holds that
\begin{align*}
  \frac{1}{2} \mathbb{E}[ M ] 
  &= \sum\limits_{i=0}^{k-1} \sum\limits_{\substack{Y \in \\ \mathcal{C}_{i} \setminus \{X_{i}, Z_{i}\}}} \frac{1}{2} \mathbb{E}[M_{Y, X_{i} \cup Z_{i}}^{i}] \\
  &\le \sum\limits_{i=0}^{k-1} \sum\limits_{\substack{Y \in \\ \mathcal{C}_{i} \setminus \{X_{i}, Z_{i}\}}} 
  \left(
    |(L_{\pi_{0}} \setminus L_{\pi_{k}^{\OPT}}) \cap L_{X_{i}, Y}| \cdot \frac{|Z_{i}|}{|X_{i}| + |Z_{i}|}
    + |(L_{\pi_{0}} \setminus L_{\pi_{k}^{\OPT}}) \cap L_{Z_{i}, Y}| \cdot \frac{|X_{i}|}{|X_{i}| + |Z_{i}|} 
  \right) \\
  & = \sum\limits_{i=0}^{k-1} \left( 
    |(L_{\pi_{0}} \setminus L_{\pi_{k}^{\OPT}}) \cap L_{X_{i}, V \setminus X_{i}}| \cdot \frac{|Z_{i}|}{|X_{i}| + |Z_{i}|}
    + |(L_{\pi_{0}} \setminus L_{\pi_{k}^{\OPT}}) \cap L_{Z_{i}, V \setminus Z_{i}}| \cdot \frac{|X_{i}|}{|X_{i}| + |Z_{i}|} 
  \right).
\end{align*}

We now rewrite the above sum in terms of pairs rather than edges:
\begin{equation}
  \frac{1}{2} \mathbb{E}[M] \le \sum\limits_{p \in L_{\pi_{0}} \setminus L_{\pi_{k}^{\OPT}}} \sum\limits_{i=0}^{k-1} \left( \frac{|Z_{i}|}{|X_{i}| + |Z_{i}|} \cdot \mathbbm{1}(p \in L_{X_{i}, V \setminus X_{i}}) 
   + \frac{|X_{i}|}{|X_{i}| + |Z_{i}|} \cdot \mathbbm{1}(p \in L_{Z_{i}, V \setminus Z_{i}})
   \right). 
  \label{ineq:global_ineq_moving_cost}
\end{equation}

  Let $p$ be a pair in $L_{\pi_{0}} \setminus L_{\pi_{k}^{\OPT}}$. $p$ has a
  strictly positive coefficient for $i$ in the above sum if and only if $p$ has
  one of its nodes contained in $X_{i}$ or $Z_{i}$. As a result, for each $i$ so
  that $p$ has a strictly positive coefficient in the sum, one of the components
  that contains a node of $p$ merges into a~bigger one in the next graph
  $G_{i+1}$.

  Let $y$ be a node in $p$, let $Y_{1} \dots Y_{N}$ be the components that
  successively merge with $y$'s component --- we call $Y_{0}$ the initial
  component containing only $y$. Summing all the coefficients corresponding to
  pair $p$ and where $y$ is in a merging component gives
  \[
    \sum\limits_{i=1}^{N} \frac{|Y_{i}|}{\sum_{j=0}^{i} |Y_{j}|},
  \]
  which is lower than $H_n$ using Lemma~\ref{lem:harm_sum_for_moving_costs}.
  There is such a sum for the two nodes of every pair $p \in L_{\pi_{0}}
  \setminus L_{\pi_{k}^{\OPT}}$ in
  inequality~(\ref{ineq:global_ineq_moving_cost}). Hence the claim holds.
\end{proof}

\begin{observation}
  \label{obs:opt_lower_bound}
  Let \OPT be an optimal offline algorithm and let $\pi_{k}^{\OPT}$ be its final permutation.
  Then, $c(\OPT) \geq |L_{\pi_{0}} \setminus L_{\pi_{k}^{\OPT}}|$.
\end{observation}

\begin{proof}
The claim directly follows from the fact that $d(\pi_{0},\pi_{k}^{\OPT}) =
|L_{\pi_{0}} \setminus L_{\pi_{k}^{\OPT}}|$.
\end{proof}

Combining Theorem \ref{th:ub_moving_cost} and Observation
\ref{obs:opt_lower_bound} and using the well-known relationships between the
harmonic sum and the natural logarithm finally proves Theorem
\ref{thm:final_comp_ratio_cliques}.

\section{An Optimal Randomized Algorithm for Lines}
\label{sec:rand_for_lines}

\subsection{Description of the Algorithm}

In this section, we present a randomized algorithm \rand for the MinLA problem
when the subgraphs $G_0, G_1, \dots G_k$ are collections of lines. We will prove
that this algorithm holds an~expected competitive ratio of $8 \ln n$ against
the oblivious adversary~\cite{borodin-book}.

Let $x_{i}$ and $z_{i}$ be the endpoints of the edge that is revealed between
$G_{i}$ and $G_{i+1}$. We call $X_{i} $ and $ Z_{i}$ the components that contain
$x_{i}$ and $z_{i}$ --- a component refers to the nodes of a connected
component. In response to $G_{i+1}$ being revealed, our algorithm \rand handles
the necessary update in two successive parts:
\begin{itemize}
  \item first, \rand places $X_{i}$ and $Z_{i}$ next to each other;
  \item second, \rand places $x_{i}$ and $z_{i}$ next to each other.
\end{itemize}

We refer to the first part as the \emph{moving} part of the update, and we refer
to the second part as the \emph{rearranging} part of the update. Next, we
describe how our algorithm performs those two parts.

\rand performs the moving part exactly like in the clique case above: $X_{i}$
moves with probability $|Z_{i}| / (|X_{i}| + |Z_{i}|)$ and $Z_{i}$ moves with
probability $|X_{i}| / (|X_{i}| + |Z_{i}|)$.

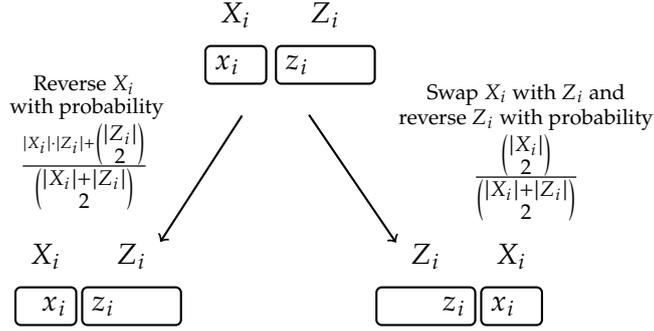
\begin{figure}[t]
  \centering
  \tikzstyle{comp}=[draw, thick, rounded corners=2pt, minimum height = 0.5cm]
  \tikzstyle{minicomp}=[draw, rounded corners=2pt, minimum height = 0.3cm]
  \begin{tikzpicture}[transform shape, scale=1]

  \def\distanceVertical{3.2cm}
  \def\distanceHorizontal{2.5cm}

  \def\widthX{0.8cm}
  \def\widthZ{1.3cm}
  \def\widthY{0.3cm}

  \def\distanceXandY{0.1cm}
  \def\distanceXandZ{0.05cm}
  \def\distanceY{0cm}

  \node[comp, minimum width=\widthX] at (0, 0 - 0*\distanceVertical) (compX1) {};
  \node (compZ1) [comp, minimum width=\widthZ, right=0cm and \distanceXandY of compX1] {};

  \node (X1) [above=0.1cm and 0cm of compX1] {$X_{i}$};
  \node (Z1) [above=0.1cm and 0cm of compZ1] {$Z_{i}$};

  \node (x1) [left=0.3cm and 0cm of compX1, anchor=west, font=\normalsize] {$x_{i}$};
  \node (z1) [left=0.3cm and 0cm of compZ1, anchor=west, font=\normalsize] {$z_{i}$};

  \node[comp, minimum width=\widthX] at (-\distanceHorizontal, 0 - 1*\distanceVertical) (compX2) {};
  \node (compZ2) [comp, minimum width=\widthZ, right=0cm and \distanceXandZ of compX2] {};

  \node (X2) [above=0.1cm and 0cm of compX2] {$X_{i}$};
  \node (Z2) [above=0.1cm and 0cm of compZ2] {$Z_{i}$};

  \node (x1) [right=0.3cm and 0cm of compX2, anchor=east, font=\normalsize] {$x_{i}$};
  \node (z1) [left=0.3cm and 0cm of compZ2, anchor=west, font=\normalsize] {$z_{i}$};

  \node[comp, minimum width=\widthZ] at (\distanceHorizontal, 0 - 1*\distanceVertical) (compZ3) {};
  \node (compX3) [comp, minimum width=\widthX, right=0cm and \distanceXandZ of compZ3] {};

  \node (X3) [above=0.1cm and 0cm of compX3] {$X_{i}$};
  \node (Z3) [above=0.1cm and 0cm of compZ3] {$Z_{i}$};

  \node (z1) [right=0.3cm and 0cm of compZ3, anchor=east, font=\normalsize] {$z_{i}$};
  \node (x1) [left=0.3cm and 0cm of compX3, anchor=west, font=\normalsize] {$x_{i}$};

  \draw [thick, shorten <=0.8cm, shorten >=0.7cm, ->] (compZ1.west) -- node [pos=0.7, anchor=south east] {\normalsize $\substack{\text{Reverse } X_{i}\\ \text{ with probability}\\ \frac{|X_{i}| \cdot |Z_{i}| + \among{|Z_{i}|}{2}}{\among{|X_{i}| + |Z_{i}|}{2}}}$} (compZ2.north);
  \draw [thick, shorten <=0.8cm, shorten >=0.7cm, ->] (compZ1.west) -- node [pos=0.75, anchor=south west] {\normalsize $\substack{\text{Swap } X_{i} \text{ with } Z_{i}\text{ and}\\ \text{reverse } Z_{i} \text{ with probability}\\ \frac{\among{|X_{i}|}{2}}{\among{|X_{i}| + |Z_{i}|}{2}}}$} (compZ3.north);

  \end{tikzpicture}

  \caption{The two possible actions of \rand to rearrange the components $X_{i}$ and $Z_{i}$ together for a particular configuration of the merging components}
  \label{fig:rearranging_possibilities}
\end{figure}

In order to perform the rearranging part of the update, our algorithm has only
two options, one for each of the two orientations of $X_{i} \cup Z_{i}$ in the
permutation. Again, \rand makes its choices by flipping a biased coin weighted
as follows: the probability to make one given choice equals the (normalized)
cost of making the other choice. For a visual description, see
Figure~\ref{fig:rearranging_possibilities}.

In the rest of this section, we will prove the following theorem.
\begin{restatable}{theorem}{thFinalCompRatioLines}
  The algorithm \rand is $8 \ln n$ competitive for the problem of online MinLA where the revealed subgraphs $G_0, \dots G_k$ are collections of lines.
  \label{thm:final_comp_ratio_lines}
\end{restatable}

\subsection{Additional Notation for the Proofs}
\label{ssec:notation}

Let $x_{i}$ and $z_{i}$ be the endpoints of the edge that is revealed between
$G_{i}$ and $G_{i+1}$ for all $0 \le i \le k-1$. Let $\mathcal{C}_{i}$ be the
set of components of $G_{i}$. We call $X_{i}, Z_{i} \in \mathcal{C}_{i}$ the
components that contain $x_{i}$ and $z_{i}$ respectively. If $T, U \in
\mathcal{C}_{i}$ are two components, we define:
\begin{itemize}
  \item $\overrightarrow{T}$ and $\overleftarrow{T}$ are the events that $T$ has one given orientation or the reversed one (an orientation of a component is the ordering of its nodes in the permutation).
    $P[\overrightarrow{T}]$ is the associated probability that \rand is in a configuration with the orientation $\overrightarrow{T}$.
  \item $L_{T,U}$ is the set of all pairs that contain exactly one node from $T$ and one node from $U$ in any order, 
    i.e., $L_{T, U} = T \times U \cup U \times T$
  \item $L_{\pi_{i}}$ is the set of all node pairs $(x, y)$ such that $x$ is on the left of $y$ in $\pi_{i}$.
  \item $L_{\overrightarrow{T}}$ is the set of all pairs $(t, t^{\prime})$ in $T$ such that $t$ is on the left of $t^{\prime}$ when $T$ has orientation $\overrightarrow{T}$.
\end{itemize}

\subsection{Upper Bound on the Moving Costs}
\label{ssec:moving_for_lines}

Calling $M$ the random variable equal to the total moving costs of \rand, the
Theorem \ref{th:ub_moving_cost} of the above clique case can also be applied and
it holds that
\[
  \mathbb{E}[M] \le 4 \ln n \cdot |L_{\pi_{0}} \setminus L_{\pi_{k}^{\OPT}}|.
\]

\subsection{Upper Bound on the Rearranging Costs}
\label{ssec:rearranging}

In this subsection, we give an upper bound of the total cost incurred by the
rearranging parts of the updates. We begin our proof by computing the
probability of a component to be one or the other orientation (given in Lemma
\ref{lem:proba_direction}). Right below (Lemma \ref{lem:to_shorten_equations})
we present an auxiliary result used to derive Lemma \ref{lem:proba_direction}.

\begin{restatable}{lemma}{lemToShortenEquations}
  \label{lem:to_shorten_equations}
  Consider $N$ real numbers $a_{1}, a_{2}, \dots, a_{N}$ and $N$ real numbers
  $b_{1}, b_{2}, \dots b_{N}$ each between $0$ and $1$, we call $\bar{b}_{i} = 1
  - b_{i}$ for all $i \in [N]$. It holds that
  \[
    \sum\limits_{t \in \{0, 1\}^{N}} \left[ \sum\limits_{i=1}^{N} t_{i} \cdot a_{i} \right] \prod\limits_{j=1}^{N} b_{j}^{t_{j}} \cdot \bar{b}_{j}^{\bar{t}_{j}} = \sum\limits_{i = 1}^{N} a_{i} \cdot b_{i}.
  \]
\end{restatable}

\begin{proof}
  \begin{align*}
  \sum\limits_{t \in \{0, 1\}^{N}} \left[ \sum\limits_{i=1}^{N} t_{i} \cdot a_{i} \right] \prod\limits_{j=1}^{N} b_{j}^{t_{j}} \cdot \bar{b}_{j}^{\bar{t}_{j}}
  & = \sum\limits_{i=1}^{N} a_{i} \cdot \left[\sum\limits_{t \in \{0, 1\}^{N}} t_{i} \cdot \prod\limits_{j=1}^{N} b_{j}^{t_{j}} \cdot \bar{b}_{j}^{\bar{t}_{j}}\right]\\
  & = \sum\limits_{i=1}^{N} a_{i} \cdot b_{i} \left[\sum\limits_{\substack{t \in \{0, 1\}^{N}\\ t_{i}=1}} \prod\limits_{\substack{j=1\\ j \neq i}}^{N} b_{j}^{t_{j}} \cdot \bar{b}_{j}^{\bar{t}_{j}}\right]\\
  & = \sum\limits_{i=1}^{N} a_{i} \cdot b_{i}
 \end{align*}
  The last equality was derived using the total probability formula.
\end{proof}

\begin{lemma}
\label{lem:proba_direction}
  Let $1 \le i \le k$.
  For any $X \in \mathcal{C}_{i}$ containing strictly more than one node, it holds that
  \[
    P[\overrightarrow{X}] = \frac{|L_{\overrightarrow{X}} \cap L_{\pi_{0}}|}{\among{|X|}{2}}.
  \]
\end{lemma}

\begin{proof}
  We prove the claim by induction on $i$. The claim clearly holds for $i = 1$.
  Let $i \in \{1 \dots k\}$, we assume that the claim holds for $i$ and prove it
  also holds for $i+1$. We distinguish between cases depending on the size of
  the merging components $X_{i}$ and $Z_{i}$.

  \begin{itemize}
  \item If both $X_{i}$ and $Z_{i}$ have size one, the claim clearly holds.
  \item If only one of $X_{i}$ or $Z_{i}$ have size one, say $X_{i}$. In the
  following, we assume that the orientation $\overrightarrow{X_{i} \cup Z_{i}}$
  refers to $X_{i}$ being on the left of $Z_{i}$ and $Z_{i}$ having the
  orientation $\overrightarrow{Z_{i}}$.
    \begin{align*}
      P[\overrightarrow{X_{i} \cup Z_{i}}] 
      &= P[X_{i} \leftof Z_{i}] \cdot P[\overrightarrow{Z_{i}}]
        + \frac{|Z_{i}|}{\among{|Z_{i}| + 1}{2}} \cdot P[X_{i} \leftof Z_{i}] \cdot P[\overleftarrow{Z_{i}}]
        + \frac{\among{|Z_{i}|}{2}}{\among{|Z_{i}| + 1}{2}} \cdot P[\overrightarrow{Z_{i}}] \cdot P[Z_{i} \leftof X_{i}]\\
      &= \frac{|Z_{i}|}{\among{|Z_{i}|+1}{2}} \cdot P[X_{i} \leftof Z_{i}]
        + \frac{\among{|Z_{i}|}{2}}{\among{|Z_{i}|+1}{2}} \cdot P[\overrightarrow{Z_{i}}]\\
      &= \frac{|X_{i} \times Z_{i} \cap L_{\pi_{0}}| + |L_{\overrightarrow{Z_{i}}} \cap L_{\pi_{0}}|}{\among{|Z_{i}|+1}{2}}\\ 
      &= \frac{|L_{\overrightarrow{X_{i} \cup Z_{i}}} \cap L_{\pi_{0}}|}{\among{|Z_{i}|+1}{2}}
    \end{align*}
    which proves the claim.
    We used the induction hypothesis and Lemma~\ref{lem:proba_left_or_right} in the above calculations.
  \item Assume now that neither $X_{i}$ nor $Z_{i}$ have size one. To prove the results,
    we use Lemma~\ref{lem:to_shorten_equations} with the following values: $N =
    3$, index $a_1$ refers to the cost to reverse $X_{i}$, $a_{2}$ is the cost
    to swap $X_{i}$ and $Z_{i}$, $a_{3}$ is the cost to reverse $Z_{i}$, $b_{1}
    = P[\overleftarrow{X_{i}}]$, $b_{2} = P[Z_{i} \leftof X_{i}]$, $b_{3} =
    P[\overleftarrow{Z_{i}}]$. With those parameters (and with the $a_{j}$s
    divided by $\among{|X_{i} + Z_{i}|}{2}$), the left term of
    Lemma~\ref{lem:to_shorten_equations} equals $P[\overrightarrow{X_{i} \cup
    Z_{i}}]$, where $\overrightarrow{X_{i} \cup Z_{i}}$ stands for the
    orientation where $\overrightarrow{X_{i}}$, $X_{i} \leftof Z_{i}$ and
    $\overrightarrow{Z_{i}}$. Using Lemma~\ref{lem:to_shorten_equations}, it
    therefore holds:
    \[
      P[\overrightarrow{X_{i} \cup Z_{i}}] = \frac{1}{\among{|X_{i}| + |Z_{i}|}{2}} \cdot 
      \left( 
        \among{|X_{i}|}{2} \cdot P[\overrightarrow{X_{i}}] 
        + |X_{i}| |Z_{i}| \cdot P[X_{i} \leftof Z_{i}] 
        + \among{|Z_{i}|}{2} \cdot P[\overrightarrow{Z_{i}}] 
      \right).
    \]

    Using the induction hypothesis and Lemma \ref{lem:proba_left_or_right}, we obtain
    \begin{align*}
      P[\overrightarrow{X_{i} \cup Z_{i}}] &= \frac{1}{\among{|X_{i}| + |Z_{i}|}{2}} \cdot 
        \left(
        |L_{\overrightarrow{X_{i}}} \cap L_{\pi_{0}}|
          + |X_{i} \times Z_{i} \cap L_{\pi_{0}}|
          + |L_{\overrightarrow{Z_{i}}} \cap L_{\pi_{0}}|
        \right)\\
      &= \frac{|L_{\overrightarrow{X_{i} \cup Z_{i}}} \cap L_{\pi_{0}}|}{\among{|X_{i}| + |Z_{i}|}{2}},
    \end{align*}
    which ends this case.
    
  \end{itemize}
  We investigated all possible three cases hence the claim also holds after $G_{i+1}$ was revealed.
  By induction, the claim holds for any $i+1$.
\end{proof}

Next, we state an auxiliary lemma that is later used to prove Lemma~\ref{lem_merge_inequality}.

\begin{restatable}{lemma}{lemGenericCostUp}
  \label{lem:generic_cost_up}
  Let be $N$ real numbers $a_{1}, a_{2}, \dots a_{N}$.
  Let be $N$ real numbers $b_{1}, b_{2}, \dots b_{N}$ each between $0$ and $1$, we call $\bar{b}_{i} = 1 - b_{i}$ for all $i \in \{1 \dots N\}$.
  It holds that
  \begin{align*}
    \sum\limits_{t \in \{0, 1\}^{N}} &\left[ \sum\limits_{i=1}^{N} \bar{t}_{i} \cdot a_{i} \right] \cdot \left[ \sum\limits_{i=1}^{N} t_{i} \cdot a_{i} \right] \prod\limits_{k=1}^{N} b_{k}^{t_{k}} \cdot \bar{b}_{k}^{\bar{t}_{k}} \le \sum\limits_{i=1}^{N} b_{i} a_{i} \left(A - a_{i}\right),
  \end{align*}
  where $A = \sum_{i=1}^{N} a_{i}$.
\end{restatable}

\begin{proof}
  \begin{align*}
    \sum\limits_{t \in \{0, 1\}^{N}} \left[ \sum\limits_{i=1}^{N} t_{i} \cdot a_{i} \right] \cdot \left[ \sum\limits_{j=1}^{N} \bar{t}_{j} \cdot a_{j} \right] \prod\limits_{k=1}^{N} b_{k}^{t_{k}} \cdot \bar{b}_{k}^{\bar{t}_{k}} 
    & = \sum\limits_{i=1}^{N} a_{i} \sum\limits_{\substack{t \in \{0, 1\}^{N}\\t_i = 1}}  \left[ \sum\limits_{j=1}^{N} \bar{t}_{j} \cdot a_{j} \right] \prod\limits_{k=1}^{N} b_{k}^{t_{k}} \cdot \bar{b}_{k}^{\bar{t}_{k}}\\
    & = \sum\limits_{i=1}^{N} a_{i} \cdot b_{i} \sum\limits_{\substack{t \in \{0, 1\}^{N}\\t_i = 1}}  \left[ \sum\limits_{j=1}^{N} \bar{t}_{j} \cdot a_{j} \right] \prod\limits_{\substack{k=1\\k \neq i}}^{N} b_{k}^{t_{k}} \cdot \bar{b}_{k}^{\bar{t}_{k}}\\
    & = \sum\limits_{i=1}^{N} a_{i} \cdot b_{i} \sum\limits_{\substack{t \in \{0, 1\}^{N}\\t_i = 1}}  \left(\left[ \sum\limits_{\substack{j=1\\j \neq i}}^{N} \bar{t}_{j} \cdot a_{j} \right] \prod\limits_{\substack{k=1\\k \neq i}}^{N} b_{k}^{t_{k}} \cdot \bar{b}_{k}^{\bar{t}_{k}}\right)\\
    & =\sum\limits_{i=1}^{N} a_{i} \cdot b_{i} \left[ \sum\limits_{\substack{j=1\\j \neq i}}^{N} a_{j} \cdot b_{j} \right]\\
    & \leq \sum\limits_{i=1}^{N} a_{i} \cdot b_{i} \left( A - a_{i}\right).
  \end{align*}
  We used Lemma~\ref{lem:to_shorten_equations} from line 4 to line 5, and used that $b_{i} \le 1$ for the last inequality.
\end{proof}

\begin{lemma}
\label{lem_merge_inequality}
Let $i \in \{0 \dots k-1\}$. Let $R_{X_{i},Z_{i}}$ be the cost of the
rearranging part of the update that follows the reveal of $G_{i+1}$. The
following holds:
\begin{align*}
  \frac{1}{2} \mathbb{E}\left[ R_{X_{i},Z_{i}} \right] 
  & \leq |(L_{\pi_{0}} \setminus L_{\pi_{k}^{\OPT}}) \cap L_{X_{i}, X_{i}}| \cdot \frac{ |X_{i}||Z_{i}| + \among{|Z_{i}|}{2}}{\among{|X_{i}| + |Z_{i}|}{2}} 
  +|(L_{\pi_{0}} \setminus L_{\pi_{k}^{\OPT}}) \cap L_{X_{i}, Z_{i}}| \cdot \frac{ \among{|X_{i}|}{2} + \among{|Z_{i}|}{2}}{\among{|X_{i}| + |Z_{i}|}{2}} \\
  & \quad\quad +|(L_{\pi_{0}} \setminus L_{\pi_{k}^{\OPT}}) \cap L_{Z_{i}, Z_{i}}| \cdot \frac{ |X_{i}||Z_{i}| + \among{|X_{i}|}{2} }{\among{|X_{i}| + |Z_{i}|}{2}}. 
\end{align*}
\end{lemma}

\begin{proof}
  In its final permutation $\pi_{k}^{\OPT}$, the optimal offline algorithm \OPT
  must have chosen one configuration for each component. In particular, \OPT
  chose one configuration for the orientation of $X_{i}$, one for the relative
  position of $X_{i}$ and $Z_{i}$, and one for the orientation of $Z_{i}$.
  Without loss of generality, we assume that \OPT configured $X_{i}$ and $Z_{i}$
  as follows: $\overleftarrow{X_{i}}$, $Z_{i} \leftof X_{i}$ and
  $\overleftarrow{Z_{i}}$. In that case, we use Lemma~\ref{lem:generic_cost_up}
  with the following values: $N = 3$, $a_{1} = \among{|X_{i}|}{2}$, $a_{2} =
  |X_{i}| \cdot |Z_{i}|$, $a_{3} = \among{|Z_{i}|}{2}$, $b_{1} =
  P[\overrightarrow{X_{i}}]$ $b_{2} = P[X_{i} \leftof Z_{i}]$ and $b_{3} =
  P[\overrightarrow{Z_{i}}]$. The obtained expression corresponds to the
  expected value of $R_{X_{i}, Z_{i}}$ if we divide it by $\among{|X_{i}| +
  |Z_{i}|}{2}/2$. We use the inequality of
  Lemma~\ref{lem:generic_cost_up} and then Lemma~\ref{lem:proba_direction} to
  replace the probabilities by their expressions. Finally, we perform the same
  reasoning as in Lemma~\ref{lem:expected_number_of_swaps} to link our upper
  bound with the final permutation of \OPT and obtain the claim.
\end{proof}

Next, we state an auxiliary lemma that is useful in the proof the next theorem.

\begin{restatable}{lemma}{lemHarmSumForReversingCosts}
  \label{lem:harm_sum_for_reversing_costs}
  Let $s_{1}, s_{2}, \dots s_{N}$ be a series of strictly positive natural numbers, let $S = \sum_{i=1}^{N} s_{i}$ be their sum.
  It holds:
  \begin{align*}
    &\sum_{i=1}^{N} \frac{s_{i}^{2}}{ \among{\sum_{j=1}^{i} s_{j}}{2}} \leq 2 H_S & \text{ and }& &\sum_{i=2}^{N} \frac{s_{i-1} \cdot s_{i}}{ \among{\sum_{j=2}^{i} s_{j}}{2}} \leq 2 H_S.
  \end{align*}
  where $H_S = 1 + \frac{1}{2} + \dots \frac{1}{S}$ is the harmonic sum until the $S$-the term.
\end{restatable}

\begin{proof}
  Proving the first inequality is direct since all the terms of its left member are smaller than those of lemma \ref{lem:harm_sum_for_moving_costs}.
  We now prove the second inequality.
  \begin{align*}
    \sum_{i=2}^{N} \frac{s_{i-1} \cdot s_{i}}{\among{\sum_{j=2}^{i} s_{j}}{2}} &\leq 2 \sum_{i=2}^{N} \frac{s_{i-1}}{\sum_{j=2}^{i-1} s_{j}} \cdot \frac{s_{i}}{\sum_{j=2}^{i} s_{j}}\\
    &\leq 2 \sqrt{\sum_{i=2}^{N} \left( \frac{s_{i-1}}{\sum_{j=2}^{i-1} s_{j}} \right)^{2}} \cdot \sqrt{\sum_{i=2}^{N} \left( \frac{s_{i}}{\sum_{j=2}^{i} s_{j}} \right)^{2} }\\
    &\leq 2 H_S
  \end{align*}
  We used the Cauchy-Schwarz inequality in the second line.
\end{proof}

\begin{theorem}
  \label{th:ub_rearranging_costs}
  Calling $M$ and $R$ respectively the moving and rearranging costs of \rand, it holds that
  \begin{align*}
    \mathbb{E}[M + R] \le 8 H_n \cdot |L_{\pi_{0}} \setminus L_{\pi_{k}^{\OPT}}|
  \end{align*}
  where $H_n$ is the harmonic sum.
\end{theorem}

\begin{proof}
  We first derive an upper bound of the expected value of $R$.
  \begin{align}
  \frac{1}{2} \mathbb{E}[R] 
    & = \sum\limits_{i=0}^{k-1} \frac{1}{2} \mathbb{E}[R_{X_{i}, Z_{i}}] \nonumber\\
    & \le \sum\limits_{i=0}^{k-1} |(L_{\pi_{0}} \setminus L_{\pi_{k}^{\OPT}}) \cap L_{X_{i}, X_{i}}| \cdot \frac{ |X_{i}||Z_{i}| + \among{|Z_{i}|}{2}}{\among{|X_{i}| + |Z_{i}|}{2}} 
    +|(L_{\pi_{0}} \setminus L_{\pi_{k}^{\OPT}}) \cap L_{X_{i}, Z_{i}}| \cdot \frac{ \among{|X_{i}|}{2} + \among{|Z_{i}|}{2}}{\among{|X_{i}| + |Z_{i}|}{2}} \nonumber \\
    & \quad\quad\quad\quad +|(L_{\pi_{0}} \setminus L_{\pi_{k}^{\OPT}}) \cap L_{Z_{i}, Z_{i}}| \cdot \frac{ |X_{i}||Z_{i}| + \among{|X_{i}|}{2} }{\among{|X_{i}| + |Z_{i}|}{2}} \nonumber \\
    & = \sum\limits_{p \in L_{\pi_{0}} \setminus L_{\pi_{k}^{\OPT}}} \sum\limits_{i=0}^{k-1} \frac{ |X_{i}||Z_{i}| + \among{|Z_{i}|}{2}}{\among{|X_{i}| + |Z_{i}|}{2}} \cdot \mathbbm{1}(p \in L_{X_{i}, X_{i}}) 
    + \frac{ \among{|X_{i}|}{2} + \among{|Z_{i}|}{2}}{\among{|X_{i}| + |Z_{i}|}{2}} \cdot \mathbbm{1}(p \in L_{X_{i}, Z_{i}}) \nonumber \\
    & \quad\quad\quad\quad + \frac{ |X_{i}||Z_{i}| + \among{|X_{i}|}{2} }{\among{|X_{i}| + |Z_{i}|}{2}} \cdot \mathbbm{1}(p \in L_{Z_{i}, Z_{i}}) \label{ineq:rearranging_costs}
  \end{align}

  Combining the upper bound (\ref{ineq:rearranging_costs}) above with the upper
  bound (\ref{ineq:global_ineq_moving_cost}) in the proof of
  Theorem~\ref{th:ub_moving_cost}, we now have upper bounds for the expected
  costs of both moving and rearranging parts of the updates, and hence for the
  total expected cost of \rand.
  
  Let $p \in L_{\pi_{0}} \setminus L_{\pi_{k}^{\OPT}}$ and let $y \in p$.  
  In the upper bounds (\ref{ineq:global_ineq_moving_cost}) and
  (\ref{ineq:rearranging_costs}), we sum and upper-bound all the positive
  coefficients associated with node $y$ and pair $p$. In case the nodes of $p$
  never belong to the same component then the upper bound of
  Theorem~\ref{th:ub_moving_cost} then the upper bound
  (\ref{ineq:global_ineq_moving_cost}) will be sufficient for that pair. We now
  assume that the nodes of $p$ eventually belong to the same component, let
  $i_{p}$ be the request index when this happens.

  Using the upper bound (\ref{ineq:global_ineq_moving_cost}) and the same
  reasoning as in Theorem~\ref{th:ub_moving_cost}, we find that the sum of the
  positive coefficients associated to $y$ and $p$ is no greater than $4
  H_{i_{p}}$ from revealed subgraphs $G_{1}$ to $G_{i_{p}}$. We now upper-bound
  the cost associated to $y$ and $p$ after $i_{p}$ using
  (\ref{ineq:rearranging_costs}). Let $Y_{i_{p}} \dots Y_{N}$ be the components
  that successively merge with $y$'s component after $i_{p}$: $Y_{i_{p}}$ is
  the component that contains the other node in $p$. We can associate the
  rearranging cost paid during the $i_{p}$'s update with the coefficient
  $\Big(\among{|Y_{i_{p}}|}{2} + \among{|Y_{i_{p}-1}|}{2}\Big) /
  \among{|Y_{i_{p}}| + |Y_{i_{p}-1}|}{2}$ in the sum. All the rearranging costs
  paid after the $i_{p}$'s update are associated with the coefficients
  $\Big(|Y_{i+1}| \cdot |Y_{i}| + \among{|Y_{i+1}|}{2}\Big) / \among{|Y_{i+1}| +
  |Y_{i|}|}{2}$. Hence, by applying Lemma~\ref{lem:harm_sum_for_reversing_costs}
  for those coefficients, we obtain that the expected cost incurred between
  request indices $i_{p}$ and $k$ is no greater than $4 (H_k - H_{i_{p}})$ for
  $y$ in the pair $p$. Applying this reasoning for the two nodes in each pair $p
  \in L_{\pi_{0}} \setminus L_{\pi_{k}^{\OPT}}$, we finally have that
\[
  \mathbb{E}[M + R] \le \sum\limits_{p \in L_{\pi_{0}} \setminus L_{\pi_{k}^{\OPT}}} 4 H_{i_{p}} + 8 (H_{k} - H_{i_{p}}) 
  \le 8 H_n \cdot |L_{\pi_{0}} \setminus L_{\pi_{k}^{\OPT}}|,
\]
which ends the proof.
\end{proof}

Finally, using the above Theorem~\ref{th:ub_rearranging_costs} with the
Observation~\ref{obs:opt_lower_bound} directly proves
Theorem~\ref{thm:final_comp_ratio_lines}.

\section{Lower Bounds}
\label{sec:lbs}

In this section, we show tightness of our analysis. First, we show a lower bound
for any randomized online algorithm, concluding that our randomized algorithm is
asymptotically optimal among all online randomized algorithms. Second, we show a
weaker lower bound for the competitiveness of the deterministic algorithms
belonging to the family of algorithms considered in Section~\ref{sec:det},
concluding that our analysis of this family is tight; however the question
whether this family is optimally competitive remains open.

\subsection{A Lower Bound of $\Omega(\log n)$ for any Randomized Algorithm}

The lower bound is an adaptation of the lower bound given by Olver et al.~for
the \emph{itinerant list update problem}~\cite{ILU}. This theorem gives an
asymptotically tight lower bound to complement
Theorems~\ref{thm:final_comp_ratio_cliques} and
\ref{thm:final_comp_ratio_lines}.

\begin{theorem}
  If a randomized online algorithm for the online minimum linear arrangement is $c$-competitive, then $c \ge \frac{1}{16}\cdot \log n$.
\end{theorem}

\begin{proof}
We apply Yao's principle to derive the lower bound. We construct a distribution
of request sequences in the following way. Let $n$ be a power of $2$ so that $n
= 2^{q}$, and $q= \log n$. First, we choose a random permutation $P$ of the $n$
nodes. Then, we construct a balanced binary tree whose leaves are the
permutation $P$ and the internal nodes are added to form the tree. Then, we
request pairs of nodes as follows. We traverse the tree level by level bottom-up
starting from the penultimate level. For each internal node $z$ of the current
level of binary tree, we issue a request between the two leaves, $u$ and $v$,
chosen by following a line in the tree from the internal node until we encounter
a leaf. Concretely, to obtain the leaf $u$, we first descend from $z$ towards
its left child once, and then we continue to descend to the right child until 
a~leaf is reached. To obtain the leaf $v$, we first descend from $z$ to its right
child once, and then we continue to descend to the left child until a leaf is
reached.

Any optimal offline algorithm pays at most $n^2$ for the constructed sequence of
requests. Consider an offline algorithm $\OFF$ that orders the leaves according to
$P$ at the beginning and then does not move. $\OFF$ produces a feasible solution
since all requests are between neighbors in $P$. Any optimal offline algorithm
pays no more than $\OFF$.  

Fix any deterministic online algorithm $\ALG$. We show that $\ALG$ pays at least
$\Omega(n^{2} \cdot \log n)$ to serve the requests. To this end, we will show
that on the request issued at each level of the binary tree, $\ALG$ pays at least
$\frac{1}{8} n^2$. Before the requests of the level $i$ arrives, $\ALG$ has 
$2^{i}$~components, each of size $2^{q-i}$. We claim that the expected cost of $\ALG$ on
the $j$-th request is $(2^{i-1} + j) \cdot 2^{2(q-i)}$. First, assume that $\ALG$
moves components only after their nodes are requested. At the $j$-th request, we
have exactly $2^{i}-2j$ components not requested in this round. As requests are
issued accordingly to a random permutation, for the $j$-th request the number of
components not yet requested at this level between the requested nodes is at
least $(2^{i}-2j)/2$. The requested components must be collocated, hence for
each of such components in between, swapping its position with the requested
component costs $(2^{q-i})^2$. In total, the cost for the $j$-th request is at
least
\[
  \sum_{j=1}^{2^{i-1}} (2^{i-1}-j)\cdot (2^{q-i})^2 = 2^{2q-3} = \frac{1}{8} n^2.
\]
Second, if $\ALG$ moves some components before they are requested, each such swap
could reduce the migration cost for the requests at the current level by at most
$2$, hence the cost of $\ALG$ for the requests at the current level is at least
$\frac{1}{16}n^2$.

Summing over the $q = \log n$ levels results in the total cost at least
$\frac{1}{16} n^2 \cdot \log n$, whereas any optimal offline algorithm pays at
most $n^2$, which concludes the proof.
\end{proof}

\subsection{A Lower Bound of $\Omega(n)$ for a Family of Deterministic Algorithms}

Now, we give a lower bound of $\Omega(n)$ for competitiveness of the family of
algorithms introduced in Section~\ref{sec:det}. Combined with
Theorem~\ref{thm:det-linear}, it implies that our analysis of these algorithms
is tight. However, the question whether the competitiveness of \emph{any}
deterministic online algorithm is linear remains unresolved.

\begin{theorem}
  \label{thm:lb-det}
  Any deterministic online algorithm that always moves to any feasible permutation with the lowest distance to $\pi_{0}$ is no better than $\Omega(n)$-competitive.
\end{theorem}

\begin{proof}
Fix any deterministic algorithm $\ALG$ that always moves to any feasible
permutation with the lowest distance to $\pi_{0}$. Consider any line topology
with an odd number of nodes, and let $x$~be a node in the middle of the line.

We construct the sequence of requests as follows. Let $y_1$ and $y_2$ be the
nodes directly on the left and right of $x$. First, we request $y_1$ and
$y_2$, and refer to $Y_2$ as the component containing $y_1$ and $y_2$. To
serve this request, $\ALG$ must place $Y_2$ either on the left or on the right of
$x$. Let $y_3$ be the neighbor of $x$ such that $x$ is between $y_3$ and $Y_2$
--- this node exists since putting $x$ at one end is not a closest
permutation to $\pi_{0}$. Then, we request $y_3$ and a node from $Y_2$.
We end up with component $Y_3 = \{y_1, y_2, y_3\}$. We continue to issue
requests growing the component $Y_i$ by issuing a request to the neighbor of $x$
that neighbors $x$ but is not contained in the component~$Y_i$. We continue this
process until we have one component of size $n-1$ and $x$ alone.

The cost of an optimal offline solution is at most $n$, since the nodes in
$Y_{n-1}$ have the same internal order as in $\pi_{0}$, so an offline algorithm
can serve all the requests by moving $x$ to either the leftmost or the rightmost
position on the line and then no further node movements are necessary.

We claim that $\ALG$ pays $\Omega( n^2 )$ for this sequence of requests. Since $\ALG$
moves to a permutation closest to $\pi_{0}$, $x$ alternates between the left and
the right side of the growing component~$Y_i$. Below we prove that this behavior
must occur. Let $L$ be the set of nodes that were on the left side of $x$ in the
initial configuration, $R$ for the nodes on the right. Let $Y_i$ be the growing
component after $i$ requests. We first show that: In $Y_i$, there are strictly
more nodes from $L$ than $R$, and hence $x$ is on the right of $Y_i$ Likewise,
In $Y_i$, there are strictly more nodes from $R$ than $L$, and hence $x$ is on
the left of $Y_i$. This is true since $\ALG$ moves to a permutation with the
smallest distance to $\pi_{0}$. Hence, any node from $L$ will stay on the left
of $x$ as long as it is not requested, and the equivalent statement holds for
$R$. Since the ordering within $Y_i$ is forced, the only ordering that ALG could
change is the order between $x$ and $S_i$, hence the result. Now, notice that:
If $x$ is on the left of $Y_i$ then the next request will add a node from $L$
into $Y_{i+1}$ If $x$ is on the right of $Y_i$ then the next request will add a
node from $R$ into $Y_{i+1}$. The nodes in $Y_{i+1}$ belong to either $L$ or
$R$, and at all times we either have a majority of nodes from $L$ or from~$R$.
The majority of nodes within $Y$ will change every second request. Each time the
majority changes, $x$ and $Y$ change their relative positions, which incurs a
final cost of $\Omega(n^2)$ for $\ALG$.
    
The cost of the optimal solution is at most $O(n)$, and the cost of $\ALG$ is
$\Omega(n^2)$, hence the competitive ratio of $\ALG$ is $\Omega(n)$.
\end{proof}

As a result of the theorems in this section, the analysis of our deterministic and the randomized algorithms is tight. Furthermore, our randomized algorithm attains asymptotically optimal competitive ratio among all randomized online algorithms.

\section{Conclusions}
\label{sec:conclusions}
This paper considered a fundamental online variant of the minimum linear
arrangement problem. The problem is motivated by the goal of understanding how
to efficiently adjust virtual network embeddings. Our main contribution is a
tight randomized online algorithms for this problem in a restricted case where
the to-be-embedded graph is either a collection of lines or a collection of
cliques. An interesting open research question is whether online algorithms
can attain the logarithmic competitive ratio also for general graphs.

\bibliographystyle{abbrv}
\bibliography{references} 

\end{document}